\documentclass[conference]{IEEEtran}
\IEEEoverridecommandlockouts
\usepackage{cite}
\usepackage{amsmath,amssymb,amsfonts}
\usepackage{algorithmic}
\usepackage{graphicx}
\usepackage{textcomp}
\usepackage{xcolor}
\usepackage{comment}
\usepackage{booktabs}
\usepackage{multirow}
\usepackage{hyperref}

\def\BibTeX{{\rm B\kern-.05em{\sc i\kern-.025em b}\kern-.08em
    T\kern-.1667em\lower.7ex\hbox{E}\kern-.125emX}}
\begin{document}

\title{Frequency-Weighted Training Losses for Phoneme-Level DNN-based Speech Enhancement}

\author{
\IEEEauthorblockN{Nasser-Eddine Monir, Paul Magron, Romain Serizel}
\IEEEauthorblockA{\textit{Université de Lorraine, CNRS, Inria, Loria} \\
Nancy, France\\
\{nasser-eddine.monir\}\{paul.magron\}@inria.fr, romain.serizel@loria.fr}
}

\maketitle

\begin{abstract}
Recent advances in deep learning have significantly improved multichannel speech enhancement algorithms, yet conventional training loss functions such as the scale-invariant signal-to-distortion ratio (SDR) may fail to preserve fine-grained spectral cues essential for phoneme intelligibility. In this work, we propose perceptually-informed variants of the SDR loss, formulated in the time-frequency domain and modulated by frequency-dependent weighting schemes. These weights are designed to emphasize time-frequency regions where speech is prominent or where the interfering noise is particularly strong. We investigate both fixed and adaptive strategies, including ANSI band-importance weights, spectral magnitude-based weighting, and dynamic weighting based on the relative amount of speech and noise. We train the FaSNet multichannel speech enhancement model using these various losses. Experimental results show that while standard metrics such as the SDR are only marginally improved, their perceptual frequency-weighted counterparts exhibit a more substantial improvement. Besides, spectral and phoneme-level analysis indicates better consonant reconstruction, which points to a better preservation of certain acoustic cues.
\end{abstract}

\begin{IEEEkeywords}
multichannel speech enhancement, binaural hearing aids, DNN-based algorithms, loss function
\end{IEEEkeywords}

\section{Introduction}
\label{sec:introduction}
Multichannel speech enhancement aims to isolate speech from unwanted noise and reverberation by leveraging spatial cues from multiple microphones. This technology is crucial in real-world applications, including hearing aids~\cite{Esra2024}, automatic speech recognition~\cite{iwamoto2022bad, zhu2022joint}, teleconferencing~\cite{rao2021interspeech}, and augmented reality~\cite{Guiraud2022spear}, where retrieving high-quality speech signals is essential for human-computer interaction.

Early speech enhancement methods used statistical techniques like beamforming and spatial filtering, exploiting assumptions about spatial properties of speech and noise~\cite{Benesty2008SpringerHandbook, Hendriks2011}. Classical approaches such as minimum variance distortionless response beamformers~\cite{Capon1969} and multichannel Wiener filters~\cite{VanDenBogaert2009} effectively suppressed interference but struggled in dynamic or reverberant settings due to spatial covariance estimation challenges. Modern deep learning techniques shifted to data-driven strategies, using deep neural networks to enhance speech~\cite{Heymann2016, Carbajal2020, fasnet, fasnet-tac, deftan2, Kimura2024, adcn}, with end-to-end models extracting spatio-temporal features from raw waveforms or spectrograms for improved robustness. More recently, transformer-based architectures such as dense frequency-time attentive network (DeFTAN-II)~\cite{deftan2} and attentive dense convolutional network~\cite{adcn} have demonstrated strong performance by modeling long-range dependencies through attention mechanisms.

Standard training loss functions such as the mean squared error or the signal-to-distortion ratio (SDR)\footnote{In this paper, all ratios are \emph{scale-invariant}~\cite{LeRoux2019sisdr}, but we omit this precision (as well as the commonly used ``SI-" acronym) for brevity.} focus on signal fidelity in a global sense, overlooking the perceptual salience of phoneme-specific spectral cues that are critical to intelligibility. For instance, consonant sounds often contain high-frequency bursts or rapid spectral transitions that are easily masked by noise but crucial for distinguishing words. Since these fine-grained structures are not emphasized by conventional losses, neural models may fail to preserve them consistently. This limitation was highlighted in our previous work~\cite{monir2024phonemescale}, where we conducted a phoneme-level evaluation of enhancement algorithms and observed that improvements in global metrics often failed to translate to better reconstruction of acoustically vulnerable phonemes. These findings motivate the development of perceptually-informed objectives that selectively emphasize spectral regions tied to intelligibility, ensuring that enhancement models not only denoise but also preserve linguistically relevant acoustic features.

To address the limitations of global-scale loss functions, we explore the use of frequency-weighted loss formulations in the time-frequency (TF) domain. Specifically, we extend the standard SDR by computing it over short-time Fourier transform (STFT) representations, allowing for fine-grained temporal and spectral resolution. We incorporate both perceptually and signal-informed frequency weighting strategies to modulate the loss according to the relative importance of different spectral regions. These include fixed weighting schemes, such as the American national standards institute (ANSI) S3.5-1997 band-importance weights~\cite{ANSI}, as well as adaptive strategies based on the relative amount of noise across frequencies.
We assess their potential as training losses for a multichannel enhancement algorithm (FaSNet). Extensive evaluation suggests that perceptually weighted SDR losses can improve frequency-weighted metrics and support better preservation of phonemes like plosives and fricatives.


\section{Methodology}
\label{sec:methodology}
\subsection{Problem statement}

Let $s$ denote a reverberant clean speech signal that is contaminated with reverberant interfering noise $n$, and $\hat{s}$ an estimate of $s$ produced by a speech enhancement network. Such a network is typically trained by minimizing a loss function that measures the discrepancy between $s$ and $\hat{s}$. A popular choice for the loss is the SDR~\cite{Vincent2006, LeRoux2019sisdr}, which was originally introduced as an evaluation metric, but later exploited as a training objective in neural speech enhancement systems~\cite{adcn, fasnet-tac}.

The SDR provides a straightforward and computationally efficient training objective for time-domain signals. However, it lacks the ability to selectively emphasize specific regions of the signal. In particular, it uniformly weights all temporal and spectral components, disregarding their perceptual significance. To alleviate this limitation, we explore reformulating SDR in the frequency or TF domains, as well as leveraging non-linear spectral range and non-uniform spectral weighting schemes.


\subsection{Computing the SDR in various domains}
\label{sec:methodology-domains}


To compute the SDR, one first decomposes the estimated signal $\hat{s}$ as follows: $\hat{s} = s_{\text{proj}} + e_{\text{dist}}$,
%
%
where $s_{\text{proj}}$ and $e_{\text{dist}}$ denote the target projection and distortion component, respectively.
The SDR is then computed as:
\begin{equation}
    \label{eq:sdr_time}
    \text{SDR} = 10 \log_{10} \left( \frac{\|s_{\text{proj}}\|^2}{\|e_{\text{dist}}\|^2} \right).
\end{equation}
To obtain a frequency domain SDR, we compute the Fourier transform of the two components, denoted $S_{\text{proj}}(f)$ and $E_{\text{dist}}(f)$, from which we derive an SDR for each frequency bin independently, defined as:
\begin{equation}
    \label{eq:sdrf}
    \text{SDR}(f) = 10 \log_{10} \left( \frac{ |S_{\text{proj}}(f)|^2 }{ |E_{\text{dist}}(f)|^2} \right).
\end{equation}
These are then averaged across all frequencies to yield the frequency SDR. Similarly, to compute a TF domain SDR, we compute the STFT of the two components, from which we derive an SDR for each TF bin $\text{SDR}(f, t)$ (defined similarly as in~\eqref{eq:sdrf} as a segmental measure), and these are then averaged across all TF bins to yield the TF SDR.

While the frequency formulation only accounts for global spectral variations, the TF formulation additionally captures rapid time-varying events—such as consonants—that are critical for phoneme intelligibility yet easily masked by noise.

Note that the distortion component $e_{\text{dist}}$ can further be decomposed into a residual interfering noise $e_{\text{interf}}$ and an artifact component $e_{\text{artif}}$. Using these instead of the distortion in the above formula allows to compute the signal-to-interference and -artifact ratios (SIR and SAR)~\cite{Vincent2006}, in either the time, frequency, or TF domains. We refer the reader to Vincent et al.'s paper~\cite{Vincent2006} for full definitions about these ratios.

\subsection{Leveraging non-linear spectral scales}

In order to account for perceptual properties in this framework, we propose grouping frequency channels according to non-linear spectral scales, that are tailored to match human auditory sensitivity. For instance, the critical bands~\cite{GLASBERG1990103} and 1/3 octave scale~\cite{ANSI} reflect the unequal contribution of frequency regions to intelligibility, as bandwidth increases with frequency~\cite{ma2009objective}. In this study, we adopt the Mel scale~\cite{Stevens1937mel}, as it offers a compact, perceptually motivated representation.

\subsection{Weighting strategies}
\label{sec:weighting_strategies}

To better reflect the perceptual or intelligibility-based relevance of different frequency bands, we propose a weighted SDR formulation, where non-negative weights $w(f, t)$ are applied to the target projection and distortion component when computing the SDR. This results in a weighted TF SDR, defined as:
\begin{equation}
    \label{eq:fwsdr}
    10 \cdot \log_{10} \left( \frac{\sum_{f, t} w(f, t) \cdot |S_{\text{proj}}(f, t)|^2}{\sum_{f, t} w(f, t) \cdot |E_{\text{dist}}(f, t)|^2} \right)
\end{equation}
This formulation allows weighting strategies based on perceptual importance to directly shape the loss landscape during optimization. Note that in~\eqref{eq:fwsdr} we average each component over TF bins prior to computing the log unlike for the frequency and TF SDRs described in Section~\ref{sec:methodology-domains}, as it was shown to yield more stable training and results in preliminary experiments. 

We consider several weighting schemes. First, we consider the standard band-importance weights defined in Table 3 of the ANSI S3.5-1997 norm~\cite{ANSI}, which specifies importance values over eighteen 1/3 octave bands. These weights were derived from intelligibility studies and reflect the relative contribution of each band to overall speech understanding in typical listening conditions.

An alternative strategy is speech-shaped weighting~\cite{ma2009objective}, which derives $w(f, t)$ from the local spectral energy of clean speech in order to favor TF regions where speech is more prominent. Specifically, we consider weights of the form ${w(f, t) \propto |S(f, t)|^\gamma}$, where $|S(f, t)|$ is the magnitude of the clean speech STFT, and we set ${\gamma = 0.2}$, as it was found to maximize correlation with perceptual quality both in Ma et al.~\cite{ma2009objective} and in our own training experiments.

Finally, we propose SIR-based weights to emphasize TF bins dominated by residual noise relative to the clean speech signal. The weights are defined as:
\begin{equation}
    \label{eq:wnotlog}
    w(f, t) = \sigma\left(-\text{SIR}(f, t)\right) = \frac{\exp\left(-\text{SIR}(f, t)\right)}{\sum_{f', t'} \exp\left(-\text{SIR}(f', t')\right)},
\end{equation}
and:
\begin{equation}
    w(f, t) = \sigma\left(-\log \text{SIR}(f, t)\right) = \frac{1 / \text{SIR}(f, t)}{\sum_{f', t'} 1 / \text{SIR}(f', t')}.
    \label{eq:wlog}
\end{equation}

Both formulations emphasize low-SIR regions by assigning them higher weights. The use of a softmax normalization ensures that the weights form a valid distribution over TF bins, making them directly comparable across different signals and scales. The logarithmic variant~\eqref{eq:wlog} introduces a smoother weighting distribution, which can mitigate sharp contrasts between bins and stabilize optimization at training.

\section{Experimental Setup}
\label{sec:experimental_setup}


\subsection{Dataset}
In our experiments, we use speech signals from the LibriSpeech dataset~\cite{librispeech}, and both speech-shaped noise (SSN, 30\% of the total noise) and ecological noise from the Disco-noise dataset~\cite{furnon2021dnn} (70\% of the total noise). SSN is generated using LibriSpeech speakers not included in the clean speech, with 50\% of the \texttt{train-clean-100} set assigned to clean speech, while the remaining 50\% is partially used to generate SSN, ensuring spectral consistency between speech and noise components.
The validation and test sets are constructed following the same strategy using the LibriSpeech \texttt{dev-clean} and \texttt{test-clean} sets, respectively.

To simulate real-world reverberation, we generate room impulse responses using the Pyroomacoustics~\cite{Scheibler2018} toolbox. The reverberation time $RT_{60}$ is sampled between 0.15 and 0.4~s. The dimensions of the room are randomly selected within physically plausible ranges (3 to 8~m length, 3 to 5~m width, and 2.5 to 3~m height), allowing for variability while maintaining realistic indoor acoustic environments. The dataset also incorporates a four-channel microphone array placed according to a binaural hearing aid setup, with inter-microphone distances corresponding to realistic placements on a human head: an interaural spacing of 0.12~m to 0.18~m, and two microphones per ear positioned with lateral and vertical offsets of 0.01~m to 0.02~m and 0.01~m to 0.015~m, respectively. The speech and noise sources are placed at random locations within the defined room. The final dataset includes mixtures generated at SIR 
levels between -10 dB and 10 dB.

\subsection{Speech enhancement algorithm}

We consider the FaSNet~\cite{fasnet-tac} algorithm due to its efficiency in real-time multichannel speech enhancement. In a nutshell, FasNet is a two-stage end-to-end time-domain beamformer. In the first stage, pre-enhancement is performed on a reference microphone by estimating time-domain beamforming filters. In the second stage, it refines the target by estimating filter coefficients using pairwise cross-channel normalized correlation features. 

We retrain the model using our dataset and the implementation from the Asteroid toolkit~\cite{Pariente2020Asteroid}, while slightly adjusting the training process to accommodate our hardware setup, and optimizing performance\footnote{The code and trained models will be provided in the camera-ready version.}. In particular, to reflect binaural hearing aid scenarios, we disable random microphone permutation to maintain a fixed reference microphone (front left). 
For efficiency, we truncate audio utterances to a maximum of 10~s, we set the patience to 15 epochs and use a batch size of 1.


\subsection{Metrics}
We evaluate speech enhancement performance using the classical time-domain SDR, SIR, and SAR~\cite{LeRoux2019sisdr}. To capture perceptual relevance across frequencies, we additionally report the frequency-weighted (FW) counterparts of these metrics as defined in Greenberg et al.~\cite{Greenberg1993}, and denoted FW-SDR, FW-SIR, and FW-SAR, respectively. The weights are computed as $|S(f, t)|^{\gamma}$, as proposed by Ma et al.~\cite{ma2009objective}, following the strategy described in Section~\ref{sec:weighting_strategies}. 
We recall that this metric does not follow the same exact definition as our proposed loss~\eqref{eq:fwsdr}, and does not use the same set of (fixed) weights, as motivated by numerical stability considerations (see Section~\ref{sec:weighting_strategies}). Nonetheless, we adopt these metrics for evaluation, given their broad use in the literature and alignment with perceptually motivated evaluation standards~\cite{Greenberg1993, ma2009objective, shi2025versa}.


To evaluate denoising performance, we report the improvement in interference reduction from input to output (i.e., before to after enhancement) using both unweighted and frequency-weighted ratios, denoted as $\text{SIR}_{in}$, $\text{SIR}_{out}$ and $\text{FW-SIR}_{in}$, $\text{FW-SIR}_{out}$, respectively. In contrast, for metrics related to artifacts and distortions, we only report the output values: $\text{SAR}_{out}$, $\text{SDR}_{out}$, $\text{FW-SAR}_{out}$, and $\text{FW-SDR}_{out}$. Finally, to quantify the impact of enhancement on speech intelligibility, we include the STOI score~\cite{STOI}, reporting both input and output values: $\text{STOI}_{in}$ and $\text{STOI}_{out}$.

\section{Results and discussion}
\label{sec:results_and_discussion}
\subsection{Utterance-level evaluation}

\begin{table*}[t]
\centering
\setlength{\tabcolsep}{2pt}
\caption{Impact of the loss domain, spectral scale, and weighting scheme onto performance.}
\begin{tabular}{lcccccccccc}
\toprule
\textbf{Loss} & \textbf{Domain} & \textbf{Scale}  & \textbf{Weighting} & $\text{SIR}_{out}$  & $\text{SAR}_{out}$  & $\text{SDR}_{out}$  & $\text{FW-SIR}_{out}$  & $\text{FW-SAR}_{out}$  & $\text{FW-SDR}_{out}$  & $\text{STOI}_{out}$  \\
\midrule
L1  & Time         & - & - & 19.3  & 8.0  & 7.3  & 8.4  & 3.6  & 5.1  & 0.75 \\
L2  & Frequency     & - & - & 18.4  & 7.9  & 7.1  & 7.8  & 3.8  & 5.0  & 0.74 \\
L3  & TF & Linear    & -  & 19.2 & 7.9 & 7.2 & 8.4 & 3.5 & 5.0 & 0.75 \\
\hline
L4  & TF & Mel    &    -  & 18.8 & 7.7 & 6.9 & 8.2 & 4.0 & 6.0 & 0.76 \\
\hline
L5 & \multirow{3}{*}{TF}  & Linear     & $|S|^{\gamma}$ & 16.5 & 7.8 & 6.6 & 6.3 & 3.4 & 3.3 & 0.74 \\
L6 & & Mel        & $|S|^{\gamma}$ & 17.4 & 7.5 & 6.6 & 7.1 & 4.0 & 5.2 & 0.76 \\
L7 & & Mel        & ANSI1997        & 16.5 & 7.1 & 6.0 & 6.5 & 4.3 & 5.9 & 0.77 \\
\hline
L8  & \multirow{4}{*}{TF} & Linear & $\sigma(-\text{SIR})$          & 20.8 & 7.3 & 6.8 & 9.7 & 3.4 & 5.1 & 0.74 \\
L9 & & Linear & $\sigma(-\log(\text{SIR}))$    & 20.2 & 7.2 & 6.6 & 9.2 & 3.4 & 5.0 & 0.73 \\
L10 & & Mel    & $\sigma(-\text{SIR})$          & 18.8 & 7.1 & 6.3 & 8.1 & 4.1 & 5.7 & 0.75 \\
L11 & & Mel    & $\sigma(-\log(\text{SIR}))$    & 19.7 & 7.7 & 7.0 & 8.9 & 4.1 & 6.1 & 0.76 \\
\bottomrule
\end{tabular}
\label{tab:results}
\end{table*}

To evaluate the impact of loss domain, spectral scale, and weighting strategies on enhancement performance, we analyze the impact of the training losses defined in Section~\ref{sec:methodology} onto performance. Table~\ref{tab:results} reports the results in terms of both standard and frequency-weighted metrics at the utterance level. For reference, the mean input metrics are $\text{SIR}_{in} = -2.3$, $\text{FW-SIR}_{in} = 0.1$, and $\text{STOI}_{in} = 0.64$. We already observe that the $\text{STOI}_{out}$ scores exhibit only minor variations across configurations.

\subsubsection{Loss domains}
First, we assess the impact of the loss domain, i.e., computing the SDR in the time (L1), frequency (L2), or TF (L3) domain, onto performance. Overall, we observe in Table~\ref{tab:results} that the three losses yield comparable performance across all metrics, with the frequency domain formulation showing slightly lower values in interference-related scores. Therefore, in what follows we consider TF domain losses, as they enable the use of frequency-dependent (as well as potentially time-dependent) weighting schemes.

\subsubsection{Spectral scales}
Here we evaluate the impact of the spectral scale used when computing the loss in the TF domain. We consider grouping frequencies using the Mel scale, which corresponds to L4 in Table~\ref{tab:results}. We observe that this Mel scale-based loss (L4) yields some improvement over using the linear scale (L3), with a particularly noticeable improvement in the $\text{FW-SAR}_{out}$ and $\text{FW-SDR}_{out}$ metrics, at the cost of a moderate drop in terms of $\text{FW-SIR}_{out}$.

\subsubsection{Speech norm and ANSI1997 weighting}

We now evaluate the impact of two frequency-based weighting strategies onto output metrics (see Section~\ref{sec:methodology}): a spectral magnitude weighting $|S|^\gamma$ that can be used either with the linear (L5) or the Mel (L6) scale, and the ANSI1997 weighting scheme when considering the Mel scale (L7). These three weighting schemes are outperformed by a baseline loss without weighting (L3) in terms of $\text{SIR}_{out}$, $\text{SAR}_{out}$, and $\text{SDR}_{out}$ as well as in terms of $\text{FW-SIR}_{out}$. However, the Mel scale variants (L6 and L7) outperform the linear scale losses (L3 and L5) in terms of $\text{FW-SAR}_{out}$ and $\text{FW-SDR}_{out}$. While the ANSI1997 weighting schemes appears to be the weakest approach in terms of classical metrics and interference reduction, it exhibits some noticeable improvement in terms of (weighted) artifact reduction compared to other schemes.

\subsubsection{SIR-based weighting}
\begin{figure}[t]
    \centering
    \includegraphics[width=0.5\textwidth]{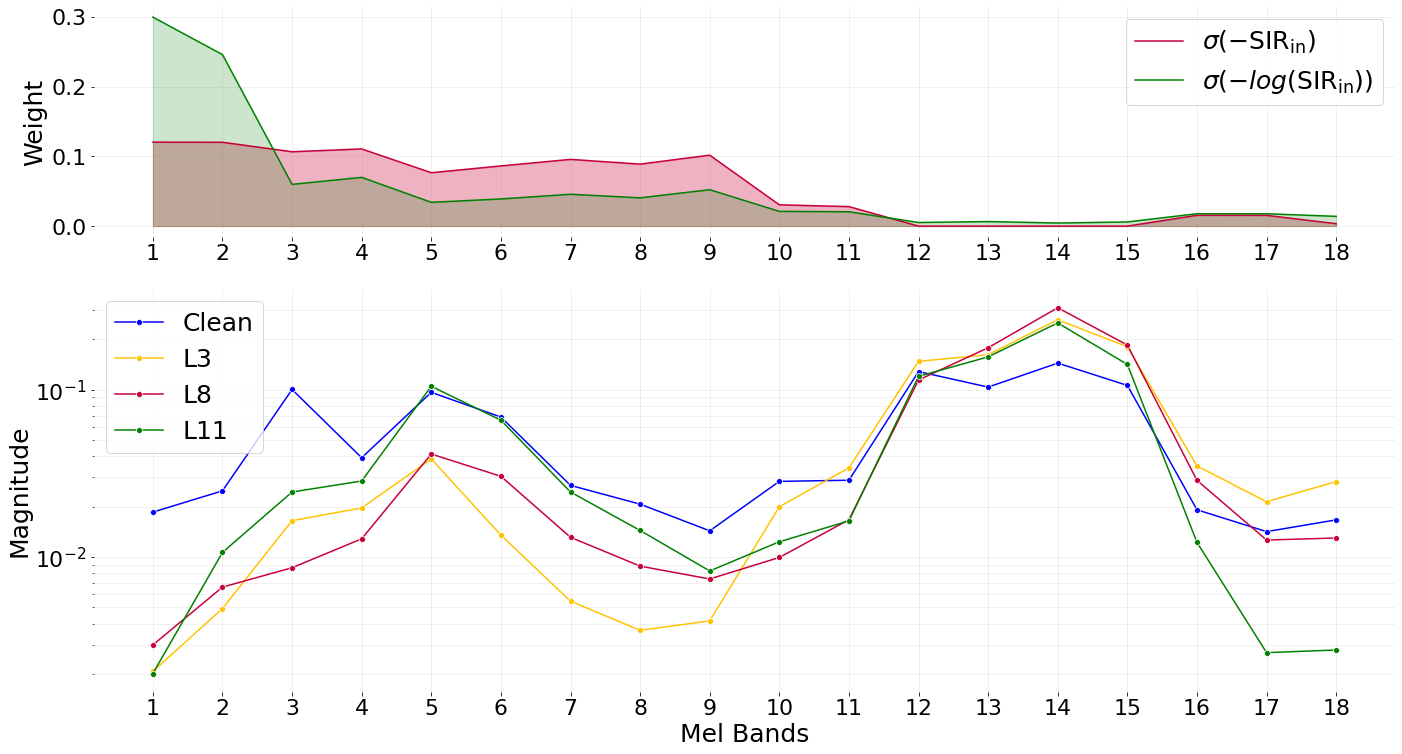} 
    \caption{Clean and enhanced (using L3, L8, and L11 losses) magnitude spectra for a female speaker utterance contaminated with 0~dB SSN (bottom), and weights used for computing L8 and L11 (top).}
    \label{fig:l3_and_l11_ssn_spectra}
\end{figure}
Finally, we investigate SIR-based weighting strategies defined in~\eqref{eq:wnotlog} and~\eqref{eq:wlog}. These approaches are applied on both linear and Mel scales (L8–L11 in Table~\ref{tab:results}). All SIR-based weighted losses exhibit a similar or slightly superior $\text{SIR}_{out}$ over no weighting (L3), but the latter outperforms these weighting schemes in terms of $\text{SAR}_{out}$ and $\text{SDR}_{out}$. SIR-based weighting improves $\text{FW-SIR}_{out}$ compared to unweighted losses, particularly when using a linear spectral scale (L8 and L9). However, using this scale comes at the cost of degraded $\text{FW-SAR}_{out}$ and $\text{FW-SDR}_{out}$. In contrast, applying SIR-based weighting on the Mel scale (L10 and L11) offers a better trade-off, preserving a high $\text{FW-SIR}_{out}$ while also improving $\text{FW-SAR}_{out}$ and $\text{FW-SDR}_{out}$. These results suggest that combining perceptually motivated spectral scales with SIR-aware weighting in the loss function can improve speech enhancement performance.

\subsection{Spectral analysis}

\begin{figure}[t]
    \centering
    \includegraphics[width=0.5\textwidth]{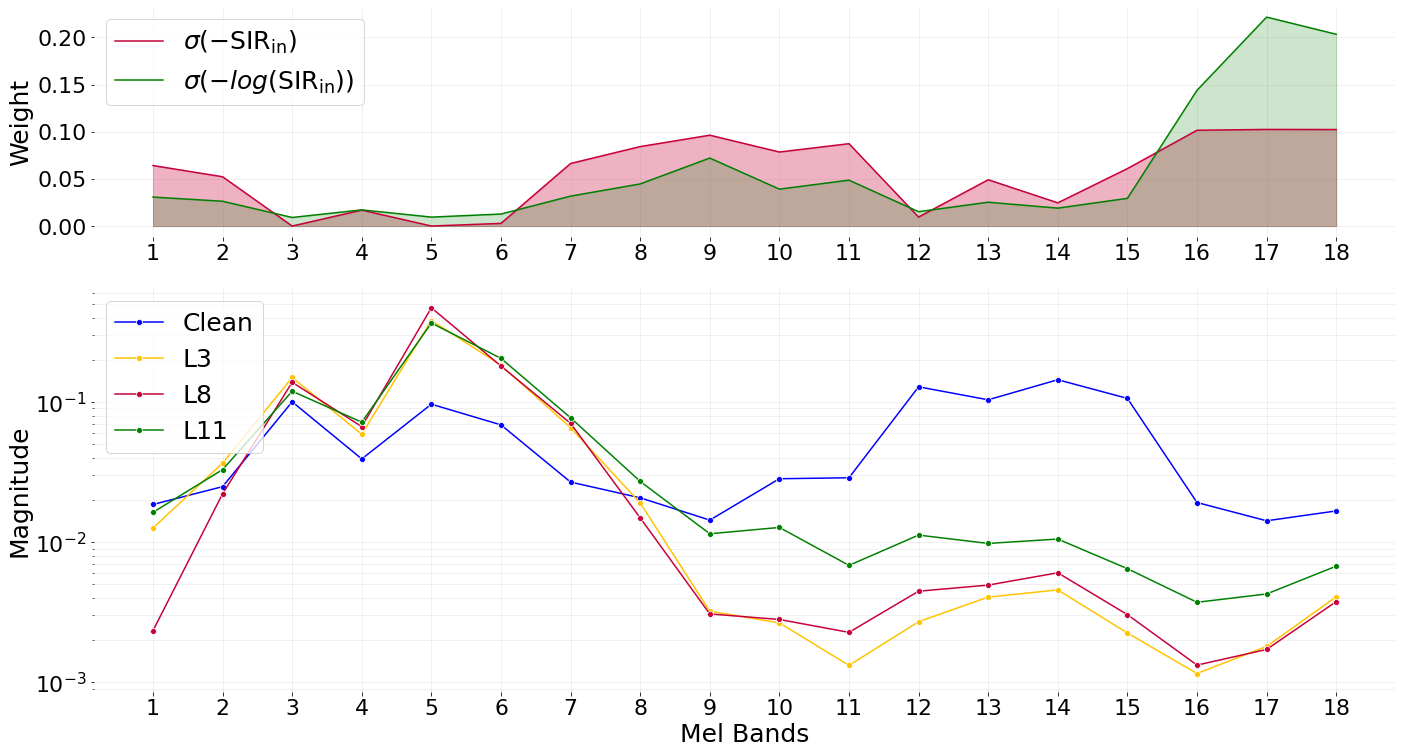} 
    \caption{Clean and enhanced (using L3, L8, and L11 losses) magnitude spectra for a female speaker utterance contaminated with 0~dB white noise (bottom), and weights used for computing L8 and L11 (top).}
    \label{fig:l3_and_l11_wn_spectra}
\end{figure}

\begin{table*}[t]
\centering
\setlength{\tabcolsep}{4pt}
\caption{Comparison of L3 and L8 losses' performance across phoneme categories.}
\begin{tabular}{lccccccccc}
\toprule
\textbf{Loss} & \textbf{Phoneme} & $\text{SIR}_{in}$ & $\text{SIR}_{out}$ & $\text{SAR}_{out}$ & $\text{SDR}_{out}$ & $\text{FW-SIR}_{in}$ & $\text{FW-SIR}_{out}$ & $\text{FW-SAR}_{out}$ & $\text{FW-SDR}_{out}$ \\
\toprule
L3 & \multirow{2}{*}{Plosive}     & \multirow{2}{*}{-6.4}  & 13.9 & 0.3 & 0.1 & \multirow{2}{*}{-9.1} & 3.5 & 1.8 & 2.6 \\
L11      & & & 15.1 & 0.3 & 0.0 & & 4.9 & 2.8 & 5.1 \\
\midrule
L3 & \multirow{2}{*}{Fricative}   & \multirow{2}{*}{-6.0}  & 14.7 & 0.7 & 0.4 & \multirow{2}{*}{-8.9} & 5.5 & 2.0 & 4.4 \\
L11      & & & 15.7 & 0.8 & 0.6 & & 6.4 & 2.9 & 5.9 \\
\midrule
L3 & \multirow{2}{*}{Approximant} & \multirow{2}{*}{0.1}  & 18.2 & 4.0 & 3.7 & \multirow{2}{*}{-4.3} & 12.6 & 3.5 & 7.9 \\
L11      & & & 18.4 & 3.9 & 3.7 & & 12.8 & 4.1 & 8.5 \\
\midrule
L3 & \multirow{2}{*}{Nasal}       & \multirow{2}{*}{-1.4}  & 19.1 & 3.7 & 3.5 & \multirow{2}{*}{-7.7} & 9.9 & 2.8 & 6.5 \\
L11      & & & 19.3 & 4.0 & 3.8 & & 10.1 & 3.9 & 7.5 \\
\midrule
L3 & \multirow{2}{*}{Vowels}      & \multirow{2}{*}{1.2}   & 17.8 & 4.0 & 3.7 & \multirow{2}{*}{-2.8} & 14.0 & 3.7 & 8.6 \\
L11      & & & 18.5 & 4.2 & 4.0 & & 13.9 & 4.1 & 8.7 \\
\bottomrule
\end{tabular}
\label{tab:phoneme_losses}
\end{table*}

Since L8 yields the highest $\text{FW-SIR}_{out}$ and L11 achieves the best $\text{FW-SAR}_{out}$ and $\text{FW-SDR}_{out}$ scores, we further investigate their spectral behavior in comparison to the baseline L3.
Figure~\ref{fig:l3_and_l11_ssn_spectra} displays clean and enhanced spectra and weights for a female speaker utterance (similar trends were consistently observed across other utterances, including those from male speakers) contaminated with SSN.

We observe that in the low-frequency range, between the 1st and 3rd Mel bands, the weights are notably high, particularly for L8, but the spectra from all three losses show noticeable deviations from the clean speech spectrum. In the mid-frequency range, specifically between the 4th and 9th Mel bands, the weights are moderately high, with L11 exhibiting slightly lower weights compared to L8. While L8 appears to follow a similar pattern to the baseline loss, L11 shows a closer alignment with the clean speech spectrum. This indicates improved spectral reconstruction and enhanced model performance in this region. Lastly, in the high-frequency range, where the weights are low, the spectra from all losses, including L8 and L11, tend to follow the same spectral trend as the baseline. However, an exception is observed for L11 in the last three Mel bands, where it diverges. Notably, in the upper bands between the 12th and 15th, none of the spectra corresponding to L3, L8, or L11 accurately follow the clean speech spectrum, reflecting a shared limitation in capturing high-frequency details.

This analysis shows that larger weights do not consistently lead to better spectral reconstruction. Nevertheless, this should be nuanced by the fact that since the SSN spectra is similar to that of the clean speech on average, the weights actually exhibit small variations over frequencies under this noise type, which might limit the impact of the weighting scheme. As a result, to better analyze the effect of the weighting schemes, we repeat the analysis with white noise, whose flat spectrum offers a more controlled setting.
Figure~\ref{fig:l3_and_l11_wn_spectra} shows the results.


We observe that at low frequencies (bands 1 to 6), the weights have very small values for both losses. In this region, the spectra for L8 and L11 progressively diverge from the clean reference as the weights decrease, suggesting a limited influence of the high SIR regions. In bands 7 to 11, where weights are higher for both weighting schemes, L8 produces spectra that are overall similar to those of the baseline L3, with only a minimal improvement in the last band. In contrast, L11 yields spectra that are visibly closer to the clean reference. A similar trend is observed in bands 16 to 18, where L8 and L3 again show comparable results despite large weight values, while L11 maintains a clear advantage. Interestingly, in bands 12 to 15, both L8 and L11 spectra deviate further from the clean target but still remain closer to it than L3, despite the very low weight values in this region. This suggests that the behavior of L8 is not clearly correlated with the magnitude of the learned weights, possibly indicating a more erratic or less effective weighting mechanism compared to L11.


These observations support the conclusion that while both L8 and L11 yield outputs that are overall more aligned with the clean spectrum than L3, the relationship between the weighting profile and spectral reconstruction is more coherent in the case of L11. This suggests that L11’s weighting function may offer finer sensitivity to frequency-dependent noise conditions, whereas the impact of L8’s weighting appears less consistent across frequency bands.


\subsection{Phoneme-level evaluation}

\begin{figure}[t]
    \centering
    \includegraphics[width=0.5\textwidth]{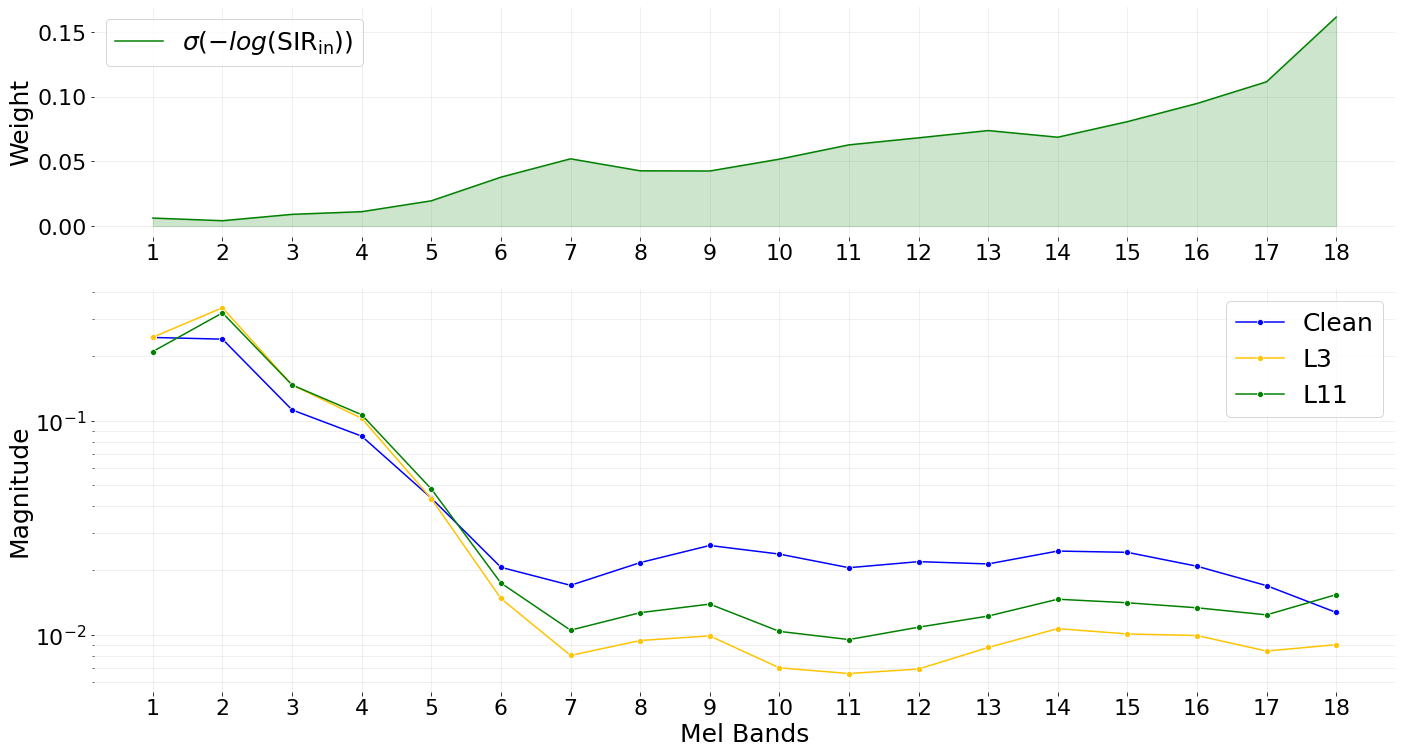} 
    \caption{Clean and enhanced (using L3 and L11 losses) magnitude spectra for plosives contaminated with 0~dB white noise (bottom), and weights used for computing the L11 losses (top).}
    \label{fig:l1_and_l11_wn_consonants}
\end{figure}

We assess the performance of the L3 and L11 loss functions across different phoneme categories. We perform segmentation into plosives, fricatives, nasals, approximants, and vowels, as in our previous study~\cite{monir2024phonemescale}. Results are detailed in Table~\ref{tab:phoneme_losses}.

This phoneme-level analysis reveals that, when considering non-weighted separation metrics such as $\text{SIR}_{out}$, $\text{SAR}_{out}$, and $\text{SDR}_{out}$, L11 does not offer clear improvements over L3. While L3 and L11 yield comparable $\text{FW-SIR}_{out}$ values on vowels, L11 demonstrates modest yet consistent gains for plosives and fricatives, which are often more challenging to enhance due to their transient or high-frequency nature. In contrast, more substantial improvements emerge in $\text{FW-SAR}_{out}$ and $\text{FW-SDR}_{out}$, where L11 surpasses L3 not only for fricatives and plosives, but also for nasals. These results suggest that the SIR-aware weighting in L11 provides a promising balance between frequency-specific artifact reduction and interfering noise suppression.

This trend is further supported by the spectral analysis of plosives contaminated with 0~dB white noise, shown in Figure~\ref{fig:l1_and_l11_wn_consonants}. Plosives were selected for their strong performance on frequency-weighted metrics, with similar patterns observed for other consonants, especially fricatives. The weights increase with frequency. In the low-frequency bands (1–5), L3 and L11 are both close to the clean speech spectrum. In the mid-frequency range (bands 6–11), both begin to deviate, but L11 remains slightly closer to the clean reference. In the high-frequency bands (12–18), L11 progressively aligns with the clean spectrum, while L3 stays relatively stable. This suggests that the weighting in L11 enhances reconstruction in noise-dominated regions.


The alignment between the perceptually motivated weight distribution and the spectral regions critical to intelligibility underscores the effectiveness of the L11 formulation. It provides a more targeted handling of interference and helps preserve spectral details that support phonemic discrimination in the presence of noise.



\section{Conclusion}
\label{sec:conclusion}
This study demonstrates that perceptually informed TF-domain SDR losses—particularly Mel-scaled, SIR-based weighting—improve the preservation of linguistically relevant spectral details, especially for plosives and fricatives. While gains in classical metrics are modest, consistent improvements are observed in frequency-weighted metrics, suggesting a better emphasis on important frequency regions. Spectral analyses further confirm enhanced reconstruction in mid-to-high frequency bands. However, limitations persist in high-frequency preservation and potential over-reliance on static weighting schemes. Future work should explore leveraging adaptive phoneme-aware weighting, incorporating auditory masking models, and evaluating generalization across varied acoustic and linguistic conditions.

\section{Acknowledgements}
This research was carried out with the support of the French National Research Agency as part of the REFINED project, “REal-time artiFicial INtelligence for hEaring aiDs” (ANR-21-CE19-0043). Experiments presented in this paper were carried out using the \href{https://www.grid5000.fr}{Grid'5000}
testbed, supported by a scientific interest group hosted by Inria and including CNRS, RENATER and several Universities as well as other organizations.

 \section{Ethical statement}
 This study uses publicly available datasets, ensuring privacy and ethical compliance by avoiding any personally identifiable information. It aims to improve accessibility for hearing-impaired users and promote inclusivity and fairness through gender-balanced dataset subsets in speech technology research.

\bibliographystyle{IEEEtran}
\bibliography{references}

\end{document}